\documentclass[showpacs,prl,superscriptaddress,,twocolumn]{revtex4}
\usepackage{amssymb}
\usepackage{amsmath}
\usepackage{graphicx}

\input{tcilatex}

\begin{document}

\title{Coherent Oscillations and Giant Edge Magnetoresistance in Singly
Connected Topological Insulators}
\date{\today }
\author{Rui-Lin Chu}
\affiliation{Department of Physics and Center for Theoretical and Computational Physics,
The University of Hong Kong, Pokfulam Road, Hong Kong, China}
\author{Jian Li}
\affiliation{Department of Theoretical Physics, University of Geneva, CH-1211 Geneva 4,
Switzerland}
\author{J. K. Jain}
\affiliation{Department of Physics, 104 Davey Lab, Pennsylvania State University,
University Park, PA 16802, USA}
\author{Shun-Qing Shen}
\affiliation{Department of Physics and Center for Theoretical and Computational Physics,
The University of Hong Kong, Pokfulam Road, Hong Kong, China}

\begin{abstract}
The topological insulators have a gap in the bulk but extended states at the
edge that can carry current. We study a geometry in which such edge states
will manifest themselves through periodic oscillations in the
magnetoconductance of a singly connected sample coupled to leads through
narrow point contacts. The oscillations occur due to quantum interference of
helical edge states of electrons traveling along the circumference of the
sample, and have a period of $\Delta B=h/eA_{\mathrm{eff}}$, where $A_{%
\mathrm{eff}}$ is the effective area enclosed by the edge states of the
sample. Our calculation indicates the possibility of a large change in the
magnetoresistance at small $B$, termed giant edge magnetoresistance, which
can have potential for application. The zero field conductance also exhibits
oscillations as a function of the Fermi energy due to interference between
edge states. The amplitude of the oscillations is governed by, and therefore
can be used to investigate, the transverse width of the edge channels.
\end{abstract}

\maketitle

Topological insulators differ from ordinary insulators in having a pair of
extended helical edge states, which results in quantum spin Hall (QSH)
effect \cite{Kane05prl,Bernevig06SCI,Konig07SCI}. Several candidates for
topological insulators have been proposed, and a non-zero conductance has
been measured experimentally in the "inverted"-band semiconductor HgTe/CdTe
quantum well in a band insulating region.\cite{Konig07SCI} More experiments
are beginning to explore the edge and surface state properties in
topological insulators.\cite%
{Hsieh08Nature,Wells09PRL,Qi09SCI,Hou09prl,Yokoyama09xxx} A direct
observation of extended edge states would be important to establish the
physics of topological insulators, and one may ask if they can also exhibit
other interesting phenomena. We demonstrate that an effective
one-dimensional ring is formed between two consecutive scatterers, which
leads to Aharonov--Bohm (AB) oscillations in conductance. An observation of
such oscillations in a singly connected geometry will constitute a direct
observation of edge transport. Such oscillations are analogous to similar
oscillations in singly connected quantum Hall systems \cite%
{Sivan89prb,Wees89prl,Jain88prl}, but with an important difference: in the
present case, the oscillations occur at very small magnetic fields. Our
results also indicate the possibility of a "giant edge magnetoresistance"
(GEMR), which is insensitive to the geometry of the device and may have
potential for practical application.

\begin{figure}[tbp]
\centering \includegraphics[width=0.45\textwidth]{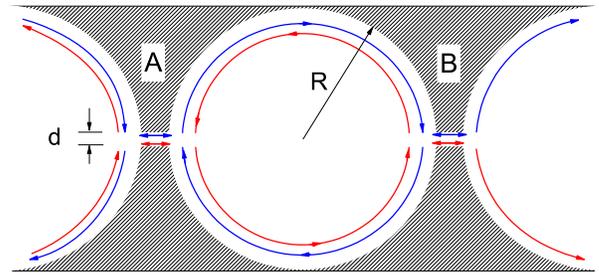} 
\caption{Schematic of the geometry studied in this manuscript, which
consists of a disk connected to two reservoirs through two quantum point
contacts. Red and blue lines indicate the chiral edge channels of spin up
and down electrons, with arrows indicating the direction of their motion. In
our calculations, we take the following parameters: width $W=500$nm, length $%
L=1030$nm; the slit width $d$ of the quantum point contacts is taken as a
variable.}
\label{fig1}
\end{figure}

We study here the device shown in FIG.\ref{fig1}, which consists of a
two-dimensional strip of a topological insulator on which two quantum point
contacts (QPCs) have been patterned thorough gates (shaded regions in FIG. %
\ref{fig1}). The QPCs define a saddle shaped confining potential, whose
height can be controlled by a gate voltage. An effective disk of area $A=\pi
R^{2}$ ($R$ is the radius of the disk) is formed in the center. An AB effect
in the device can be expected intuitively because a topological insulator
possesses a pair of independent gapless edge states of different spins
moving in opposite directions, each forming an ideal one-dimensional loop
around the disk. The two edge states are independent because no
backscattering is allowed at a given sample edge even in the presence of
weak time-reversal invariant disorder. We note here that spin is not a good
quantum number in topological insulators because of spin orbit coupling. In
the absence of a magnetic field, the actual edge states are eigenstates of
the time reversal operator; their characterization as spin up and down is
not precisely correct, and the word \textquotedblleft spin" below is to be
viewed more generally as the quantum number denoting the two states of a
Kramers doublet. The effect of magnetic field, which breaks time reversal
symmetry, is complicated. In our considerations below we assume that the
magnetic field is sufficiently weak that the Zeeman term can be neglected;
in that limit, the field does not open any gap in the spectra for the edge
states, as discussed in detail in Ref. \cite{Zhou08prl}.

In the remainder of the article we study this geometry quantitatively. We
will assume that the weak magnetic field $\mathbf{B}$ is normal to the
plane. Following Ref. \cite{Sivan89prb} we consider a spin-up (or spin-down)
electron travelling from the left hand side (LHS) in FIG. 1. At the LHS\
junction it splits into two partial waves: one is transmitted through the
QPC into the disk with amplitude $t$, and the other is transmitted across
the QPC with an amplitude $r$ causing a backscattering. We denote the wave
function amplitudes in the upper edge and lower edge, right after the LHS
junction, by $u_{1}$ and $d_{1}$, respectively. The corresponding amplitudes
in the vicinity of RHS junction are $u_{2}=u_{1}\exp [i\theta /2]$ and $%
d_{2}=d_{1}\exp [-i\theta /2]$, where $\theta =2\pi \phi /\phi _{0}+\delta
\theta $, $\phi _{0}=h/e$ a fundamental unit of magnetic flux, $\phi =\pi R_{%
\mathrm{eff}}^{2}B$ is the magnetic flux threading the effective
one-dimensional loop with an effective radius $R_{\mathrm{eff}}$, and $%
\delta \theta $ is the phase acquired by the wave function traveling along
the loop $\delta \theta =2\pi kR_{\mathrm{eff}}$. A partial wave goes
through the RHS\ slit with an amplitude $t^{\prime }$ and across the slit
with an amplitude $r^{\prime }$. To simplify the problem, we assume
identical reflection and transmission amplitudes for the two slits, $%
t=t^{\prime }$ and $r=r^{\prime }$. Using the theory of multi-scattering
processes \cite{Datta} it then follows that the total transmission for
spin-up electron through the slit A and B is given by 
\begin{equation}
T^{\uparrow }(B)=\frac{T_{0}{}^{2}}{T_{0}{}^{2}+4(1-T_{0})\cos ^{2}\frac{%
\theta +\theta _{0}}{2}}.  \label{eq:Tup}
\end{equation}%
Here $T_{0}=\left\vert t^{2}\right\vert $ is the transmission coefficient of
an electron through a slit, and $\theta _{0}=\arg (rr^{\prime }).$ Resonant
tunneling occurs for $\cos \frac{\theta +\theta _{0}}{2}=0$, i.e., $%
T^{\uparrow }(B)=1$ for any $T_{0}$. The transmission coefficient for a
spin-down electron $T^{\downarrow }(B),$ which is the time reversal
counterpart of spin-up electron at $-B$ field, is given by $T^{\downarrow
}(B)=T^{\uparrow }(-B)$. According to the Landauer-Buttiker formula,\cite%
{Landauer,Buttiker} the total conductance is 
\begin{equation}
G(B)=\frac{e^{2}}{h}[T^{\uparrow }(B)+T^{\downarrow }(B)].
\end{equation}%
The AB oscillations in the conductance $G$ as a function of the magnetic
flux $\phi $ through the disk are therefore expected to be symmetric with
respect to the direction of the magnetic field.

To make further progress, we undertake a numerical calculation for the
specific case of an \textquotedblleft inverted"-band HgTe/CdTe
heterojunction. We consider with the effective Hamiltonian for HgTe/CdTe
quantum well\cite{Bernevig06SCI} 
\begin{equation}
\mathcal{H}=\left( 
\begin{array}{cc}
h(k) & 0 \\ 
0 & h^{\ast }(-k)%
\end{array}%
\right) ,  \label{eq:ham}
\end{equation}%
where, in the spin-up sector, $h(k)=-Dk^{2}+A(k_{x}\sigma _{x}+k_{y}\sigma
_{y})+(M-Bk^{2})\sigma _{z},$ and $k=(k_{x},k_{y})$ is the wave vector in
two dimension, $k$ is the Pauli matrix vector. $A$, $B$, $D$ and $M$ are
sample specific parameters, and are functions of the thickness of the
quantum well; here we take $A=364.5$meV nm$^{2}$, $B=-686$meVnm$^{2}$, $%
D=-512$meVnm$^{2}$, $M=-10$meV. In the spin-down sector $h^{\ast }(-k)$ is
the time reversal counterpart of $h(k)$. The existence of edge states in
this model has been discussed previously.\cite{Konig08JSPS,Zhou08prl}

To determine the conductance through the disk subjected to a perpendicular
and weak field $B$, we use the Keldysh Green function technique to calculate
the transmission coefficients $T$ numerically in the Landauer-B\"{u}ttiker
formalism\cite{Datta}. For this purpose, we use a tight-binding model that
reproduces the Hamiltonian in Eq. (3) in the continuum limit, and include
the magnetic field through the substitution $\mathbf{k}\rightarrow \mathbf{k}%
-\frac{e}{\hbar }\mathbf{A}$ ($\mathbf{A}$ is the vector potential). In the
present work, more $400\times 200$ lattice sites and the lattice space $%
a=2.5 $nm were used in the calculation, and the lattice size effects are
vanishingly small, so all results reported here reflect the continuum limit.

\begin{figure}[tp]
\includegraphics[width=0.45\textwidth]{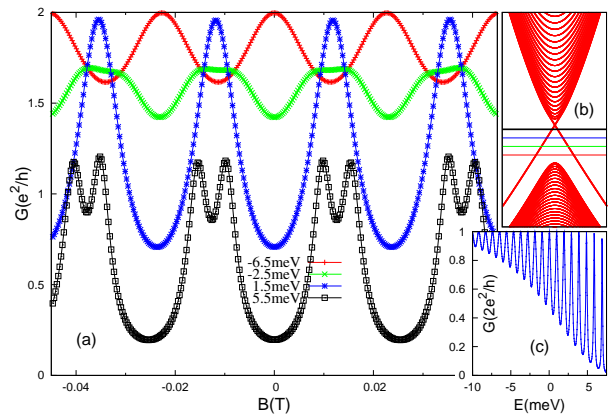} 
\caption{(a) Magnetoconductance $G$ for several energies in the band gap.
(b) The locations of the energies for the curves in panel (a). (c)
Conductance $G^{\uparrow }$ as a function of the Fermi energy at zero
magnetic field ($\mathbf{B}=0$). All calculations assume point contact
constriction width of $d=40$nm. }
\label{fig2}
\end{figure}

FIG. 2 shows the conductance for several Fermi energies, whose locations in
the band structure are illustrated in FIG. 2(b). The conductance exhibits
periodical oscillations in $B$, with the period determined by the magnetic
flux. For most parameters, the maximal value of the conductance does not
reach $2e^{2}/h$ because $T^{\uparrow }(B)$ and $T^{\downarrow }(B)$ in
general do not satisfy the resonance condition simultaneously. A new feature
is the rapid oscillations in the conductance at $B=0$ as a function the
Fermi energy $E_{f}$, as shown in FIG. 2(c). These oscillations are caused
by the phase shift $2\pi kR_{\mathrm{eff}}$ where the dispersion of the edge
state $E(k)\approx -MD/B\pm A\sqrt{1-D^{2}/B^{2}}k\approx +7.46\pm 242.5k$
meV ($k$ in unit of nm$^{-1}$) is linear in $k$. The increasing oscillation
amplitude reflects that the spatial distribution of edge state is dependent
on the Fermi energy. Accordng to Zhou \emph{et al.}\cite{Zhou08prl} the
spatial distribution of the edge state near the boundary has the form $\phi
(y)=(e^{-y/\xi _{1}}-e^{-y/\xi _{2}})/\sqrt{\frac{\xi _{1}}{2}+\frac{\xi _{2}%
}{2}-\frac{2\xi _{1}\xi _{2}}{\xi _{1}+\xi _{2}}}$ where $\xi _{1}\approx
317/(240+E)$ and $\xi _{2}\approx 317/(13.4-E)$ nm ($E$ in units of meV)
near the crossing point $k=0$ for the parameters adopted in this paper.
Because we typically have $\xi _{1}<<\xi _{2}$, the spatial distribution of
the edge state is determined dominantly by $\xi _{2}$. Thus the edge state
becomes wider with increasing energy, ultimately evolving into a bulk state
when $E\geq 13.4$meV. Correspondingly, the transmission coefficients
decreases and the oscillation amplitude increases according to Eq.(1).

\begin{figure}[tp]
\includegraphics[width=0.45\textwidth]{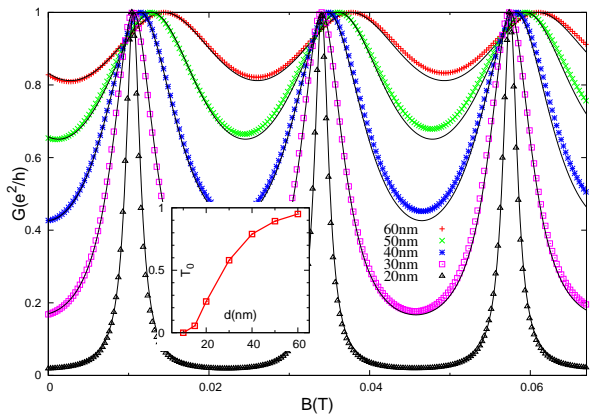} 
\caption{The magnetoconductance of a single spin channel for several widths $%
d$ of the quantum point contacts, obtained numerically for the sample of
Fig. 1. The black curves are fits using Eq. (\protect\ref{eq:Tup}). Smaller $%
d$ produces smaller $t$ (CHECK) and hence larger oscillation amplitude. The
insert shows the relation between $T_{0}$ and $d$. The period is independent
of $d$. All curves are evaluated for $E_{f}=0.5$meV. }
\label{fig3}
\end{figure}

FIG. 3 shows the dependence of oscillations on the slit width $d$. The solid
lines are the best fits from Eq.(1) with three adjustable parameters $T_{0}$%
, $R_{\mathrm{eff}}$ and $\delta \theta .$ The effective radius $R_{\mathrm{%
eff}}=\sqrt{2\phi _{0}/B_{0}}$ can be deduced from the period $B_{0}$ of the
oscillation; its ratio to the radius of the disk is $R_{\mathrm{eff}%
}/R\approx 0.95,$ which reflects the finite width of an edge state near the
boundary of the disk. The inset in FIG. 3 shows how $T_{0}$ depends on the
slit width $d$. As expected, $T_{0}$ vanishes for $d/\xi _{2}<<1$ and
approaches unity for $d/\xi _{2}>>1$. Conductance oscillations are
suppressed in both limits; in particular, for $d/\xi _{2}>>1$ it has the
universal value $2e^{2}/h$ \cite{Kane05prl,Bernevig06SCI} and displays no
magnetoresistance. The AB oscillations occur only when the slit width is
comparable to the spatial distribution of edge state. The numerical results
are in good agreement with Eq.(1) in the first one or two periods, but
exhibit an increasing deviation for a larger field $B$, which may originate
from the neglect, in our fits, of the dependence of the energy dispersion of
the edge states on magnetic field (we use the $B=0$ dispersion when fitting
the numerical data).

\begin{figure}[tp]
\includegraphics[width=0.45\textwidth]{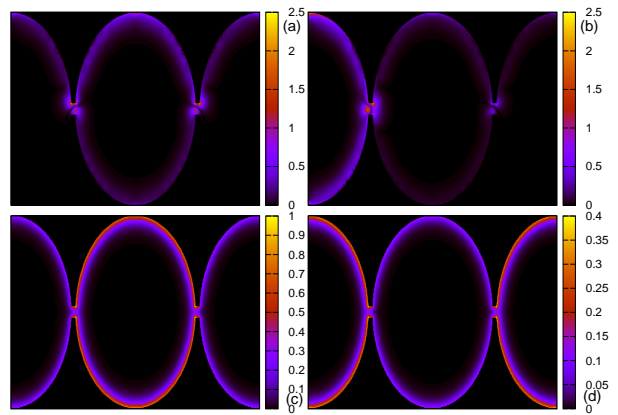} 
\caption{Panels (a) and (b) show the local current density distributions on
the sample of Fig.1 for magnetic fields $\mathbf{B}$ at a peak and a valley
of the magnetoresistance, which correspond to maximum transmission and
maximum reflection, respectively. The current reflection is totally
suppressed in (a). The panels (c) and (d) show the corresponding local
density of states distributions. The calculation assumes the parameters $%
d=30 $nm and $E=0.5$meV.}
\label{fig4}
\end{figure}

The quasi-one dimensional quantum interference behavior can be further
illustrated by plotting the spatial profile of local current and the local
density of states in a nonequilibrium situation\cite{Li07prb}, where Fermi
surface is slightly higher in the left lead so that the electrons flow from
left to right. Current flow patterns for the up spin sector are shown in
Figs. 4(a) and 4(b). The current flows only along the boundary of the device
with definite chirality (see FIG.\ref{fig1}). At the resonance transmission $%
T_{0}=1$, the backward current reflection at QPC A is totally suppressed,
which corresponds to a peak in the magnetoresistance oscillations. The
maximum reflection is observed at the valley of the oscillations. When a
resonance transmission occurs, the maximum of local density of states is
found in the central disk edge while, at minimum transmission, the maximum
local density of states is observed in the terminals.

\begin{figure}[tp]
\includegraphics[width=0.45\textwidth]{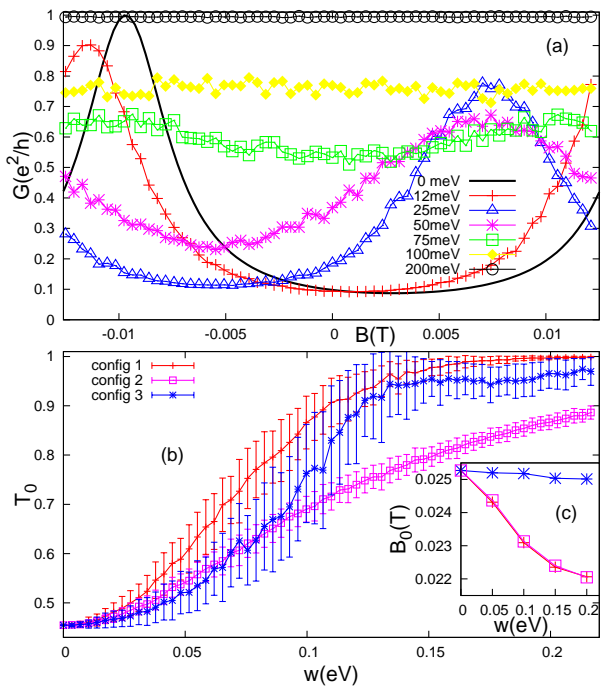}
\caption{(a). Conductance oscillations in the spin up sector as a function
of disorder. The oscillations gradually flatten but electron tunnelling
through the slit is significantly enhanced with increasing disorder
strength. (b) The disorder dependence of transmission coefficient through a
single QPC $T_{0}$ for three different spatial distributions of disorders.
Red line - disorder is distributed over the entire device. Blue line -
disorder is distributed only in the vicinity of the QPC slit. Pink line -no
disoder in the vicinity of the QPC slit. The inset in (b) shows the variance
of the period of magnetic field as a function of the disorder strength for
two different spatial distribution of disorder. Averaging is performed over
200 disorder samples; the parameters are chosen as $d=40$nm, $E=5.5$meV.}
\label{fig5}
\end{figure}

The feasibility of the observations of the predicted magnetoresistance
oscillations depends in part on how sensitive their amplitude is to the ever
present disorder. Given that time reversal invariant disorder does not cause
any backscattering between two spin channels at the same edge\cite%
{Kane05prl,Wu06prl,Xu06prb}, one might expect that it also does not affect
the coherent magnetoresistance oscillations. We study the effect of disorder
by introducing disorder as a random on-site energy with a uniform
distribution within $[-w/2,w/2]$. The results, displayed in FIG. 5(a),
demonstrate that disorder diminishes the amplitude of oscillations while
enhancing the average conductance. The reduction of the coherent
oscillations originates from an impurity configuration dependent phase shift 
$\delta \theta $ in Eq. (1); the disorder averaging over the phase shift
smears the quantum interference, completely suppressing it when the variance
of $\delta \theta $ is comparable with $\pi $. It is noteworthy that the AB
oscillations survive up to quite large disorder. For comparison, the hopping
matrix element in Eq.(3) is $t=-D/a^{2}\simeq 82$ meV, and numerical results
indicate that the oscillation is almost suppressed when the on-site disorder 
$\omega \geq t$. The enhancement of the average conductance is surprising at
first sight (see FIG. 5(b)), but we attribute it to a squeezing the spatial
distribution of the edge states by disorder. Such squeezing can also be seen
in the slight decreasing of the oscillation period of the conductance with
large disorder strength as the effective radius $R_{eff}=\sqrt{2\phi
_{0}/B_{0}}$ (see the insert in FIG. 5(b).). Also the disorder around the
slit can increase the tunneling probability as the higher disorder can close
the energy gap (caused by the finite size of the slit) near the Fermi
surface \cite{Beverly02PNAS}. Of course, even stronger disorder leads to the
Anderson localization, and the edge states will be destroyed \cite%
{Li09prl,Onoda07prl,Sheng06prl}, resulting in a lack of conductance.

All our preceding discussion assumes strictly zero temperature. Scattering
between the spin up and spin down edges becomes possible at nonzero
temperatures due to inelastic scattering involving phonons, which is
believed to cause the deviation from perfect quantization in the experiments
of Ref. \cite{Konig07SCI}. The AB oscillations discussed above will be
observable only for disks with perimeters less than the phase coherence
length of the edge states.

In conclusion, we have studied theoretically a geometry in which the edge
states of topological insulators are predicted to produce coherent
oscillations in the magnetoconductance due to Aharonov-Bohm interference.
This physics also produces a giant edge magnetoresistance in a weak field;
FIG. 2 shows that the conductance may change by an order of magnitude for a
tiny field $B\approx 0.01$T. If confirmed by experiments, this may be of
practical interest. The giant edge magnetoresistance can be controlled by
varying the slit width $d$ and the size of the disk. We have also studied
its sensitivity to disorder, and found that it persists for quite large
disorder. The amplitude of AB oscillations also contains information about
the transverse spatial extent of the edge channels.

This work was supported by the Research Grant Council of Hong Kong under
Grant No.: HKU 703708, and HKU 10/CRF/08.

\end{document}